\documentclass[11pt]{article}
\usepackage{latexsym}
\usepackage{graphicx}
\usepackage{epstopdf}
	\DeclareGraphicsExtensions{.eps}
\usepackage{multicol}
\usepackage{caption}
\usepackage{amssymb}
\usepackage{amsmath}
\newtheorem{thm}{Theorem}[section]

\newtheorem{lem}{Lemma}[section]

\newtheorem{defn}{Definition}[section]

\newtheorem{exam}{Example}[section]

\numberwithin{equation}{section}

\newcommand{\E}{\mbox{E}}

\begin{document}

\vspace*{.5cm}
\begin{center}


{\LARGE{\textbf{ Weighted cumulative entropies: An extension of CRE and CE}}}
\bigskip

{\Large{\bf{Yuri Suhov\footnote{Math Dept, Penn State University,
PA 16802, USA; DPMMS, University of Cambridge, CB30WB, UK; IITP,
RAS, 127994 Moscow GSP-4, Russia. E-mail: ims14@ps.edu, yms@statslab.cam.ac.uk},\;\; Salimeh Yasaei Sekeh\footnote{Department\;of\;Statistics,\;Federal\;University\;of\;S$\tilde{\rm a}$o\;Carlos\;(UFSCar),\;S$\tilde{\rm a}$o\;Carlos,\;Brazil.\;E-mail: sa$_{-}$yasaei@yahoo.com}}}}

\bigskip



\end{center}
\bigskip


\vspace{0.5cm}

\mbox{\quad}
\def\fB{\mathfrak B}\def\fM{\mathfrak M}\def\fX{\mathfrak X}
 \def\cB{\mathcal B}\def\cM{\mathcal M}\def\cX{\mathcal X}
\def\bu{\mathbf u}\def\bv{\mathbf v}\def\bx{\mathbf x}\def\by{\mathbf y}
\def\om{\omega} \def\Om{\Omega}
\def\bbP{\mathbb P} \def\hw{h^{\rm w}} \def\hwphi{{h^{\rm w}_\phi}} \def\bbR{\mathbb R}
\def\beq{\begin{eqnarray}} \def\eeq{\end{eqnarray}}
\def\beqq{\begin{eqnarray*}} \def\eeqq{\end{eqnarray*}}
\def\rd{{\rm d}} \def\Dwphi{{D^{\rm w}_\phi}}
\def\BX{\mathbf{X}}
\def\mwe{{D^{\rm w}_\phi}}
\def\DwPhi{{D^{\rm w}_\Phi}} \def\iw{i^{\rm w}_{\phi}}
\def\bE{\mathbb{E}}
\def\1{{\mathbf 1}} \def\fB{{\mathfrak B}}  \def\fM{{\mathfrak M}}
\def\diy{\displaystyle} \def\bbE{{\mathbb E}} \def\bbP{{\mathbb P}}
\def\bF{\bar{F}}
\def\ew{{\mathcal{E}^{\rm w}_{\phi}}}
\def\cew{{\overline{\mathcal{E}}^{\rm w}_{\phi}}}
\def\lam{\lambda}

{\bf Abstract.} We generalize the weighted cumulative entropies (WCRE and WCE), introduced in \cite{M}, for a system or component lifetime. Representing properties of cumulative entropies, several bounds and inequalities for the WCRE is proposed. 


\section{Introduction. The weighted cumulative entropies}

 An important measure of the uncertainty is entropy, commonly termed the Shannon information measure, \cite{Sh}. The current paper deals with {\it weighted} entropy; for the definition and initial results on weighted entropy the reader is referred to \cite{BG,G}.

Furthermore, the entropy of the residual lifetime $X_t=[X-t|X>t]$ as a dynamic measure of uncertainty was considered in \cite{E}. Recently, further progress as introducing cumulative entropy, cumulative residual entropy and  weighted cumulative entropy was made in \cite{RCVW,CL,M}.

The purpose of this work is to obtain a number of results for weighted cumulative entropies in the case where the weight is a general non-negative function.

Let $X$ be a non-negative absolutely continuous random variable describing a component failure time, with the probability density function (PDF), $f(x)$, the cumulative distribution function (CDF), $F(x)=P(X\leq x)$, and the survival function (SF), $\bF(x)=P(X>x)$.
\begin{defn}
 Given a function $x\in \bbR\mapsto\phi (x )\geq 0$, and an RV
$X:\;\Om\to\bbR$, with a PDF $f$,
the {\bf weighted cumulative residual entropy} (WCRE) of $X$ (or $F$) with  weight function (WF) $\phi$ is defined by
\beq\label{eq:WCRE}
\ew (X)=\ew (F) =-\int_{\bbR_+}\phi (x )\bbP(|X|>x)\log\,\bbP(|X|>x)\rd x. \eeq
Note that a standard agreement $0=0\cdot\log\,0=0\cdot\log\,\infty$ is adopted throughout the paper.\\
Given the CDF, $\bx\in\bbR_+ \mapsto F(\bx )\in [0,1]$, with WF $\phi$, the {\bf weighted cumulative entropy} (WCE) of non-negative random lifetime $X$ is defined by
\beq\label{eq:WCE}
\cew (X)=\cew (F) =-\int_{\bbR_+}\phi (x )\bbP(|X|\leq x)\log\,\bbP(|X|\leq x)\rd x. \eeq
Particularly when $\phi(x)=x$ the WCRE and WCE in (\ref{eq:WCRE}) and (\ref{eq:WCE}) can be turned out as (8) and (9) in \cite{M}. In what follows, we intend to use the same abbreviation as in \cite{M} for the weighted cumulative residual and weighted cumulative entropies.
\end{defn}
\vskip .5 truecm
\begin{exam}
{\rm{(WCRE of the uniform distribution)}} Consider an RV $X$ with uniform distribution in the interval $a<b$, $f(x)=\diy \frac{1}{(b-a)}$. Then the WCRE is the following:
\beqq \begin{array}{ccl}\ew(F)&=&-\diy\int_a^b \phi(x) (1-\frac{x}{b-a})\log (1-\frac{x}{b-a}) \rd x\\
&=& \diy (b-a)\int_{\frac{b-2a}{b-a}}^{\frac{-a}{b-a}} \phi((b-a)(1-y))y \log y \rd y.
\end{array}\eeqq
In particular, with $\phi(x)=x$, one obtains:
\beqq \begin{array}{l}\ew(F)=\diy\frac{1}{36(a-b)}\bigg[24 a^3\log\left(\frac{2a-b}{a-b}\right)-15a^2b-a^3-5b^3+6\log\left(\frac{2a-b}{a-b}\right)b^3\\
\qquad\diy -18 ab^2\log\left(\frac{2a-b}{a-b}\right)+21ab^2+6a^3\log\left(\frac{a}{a-b}\right)-18a^2b\log\left(\frac{a}{a-b}\right)\bigg].\end{array}\eeqq
\end{exam}
\vskip .5 truecm
\begin{exam}
{\rm{(WCRE of the Gaussian distribution)}} Let $g(x)$ be the Gaussian PDF with mean $\mu$ and variance $\sigma^2$. Therefore, the SF is obtained as $\bar{G}(x)=erfc(\frac{x-\mu}{\sigma})$. Here $erfc(x)$ is the complementary error function:
\beqq erfc(x)=\frac{1}{\sqrt{2\pi}}\int_x^\infty \exp(-\frac{t^2}{2})\rd t.\eeqq
In accordance with (\ref{eq:WCRE}) we obtain:
\beqq \ew(G)=-\int_0^\infty\phi(x)erfc(\frac{x-\mu}{\sigma})\log erfc(\frac{x-\mu}{\sigma}) \rd x.\eeqq
\end{exam}
\vskip .5 truecm

Given an RV $X$ with CDF $F(x)$ and SF $\bF(x)$, set,
\beqq m^{\rm w}_F(t)=\frac{1}{\bF(t)}\int_t^\infty \phi(x)\bF(x)\rd x\quad \textrm{and} \quad \overline{m}^{\rm w}_F(t)=\frac{1}{F(t)}\int_0^t \phi(x) F(x) \rd x.\eeqq
Pictorially $m^{\rm w}_F(t)$ represents the weighted mean inactivity time (WMIT) and then $\overline{m}^{\rm w}_F(t)$ the weighted mean residual time (WMRT).
\begin{lem}
{\rm{(Cf. Proposition 2.1 from \cite{M}.)}} Let $X$ be an absolutely continuous RV. Then
\beq \ew(F)=\bbE\big(m^{\rm w}_F(X)\big),\eeq
and
\beq \cew(F)=\bbE\big(\overline{m}^{\rm w}_F(X)\big).\eeq
\end{lem}

{\bf Proof.}\;The proof follows directly with the same methodology in \cite{M} but replacing $\phi(x)$ in $x$.\\
\def\bG{\bar{G}}
\begin{defn}
Given two functions $x\in \bbR^+\mapsto \bF(x )\in[0,1]$ and $x\in \bbR^+\mapsto \bG(x )\in[0,1]$, the {\bf relative} WCRE of $\bG$ relative to $\bF$ for given WF $\phi$ is defined by
\beq \label{eq:4.01} D^{\rm w}_{\phi}(\bF\|\bG)=\diy\int\limits_{\bbR_+} \phi(x) \bF(x)\log\;\frac{\bF(x)}{\bG(x)} \rd x. \eeq
Alternatively, $D^{\rm w}_{\phi}(\bF\|\bG)$ can be termed as weighted Kullback Leibler divergence between SFs $\bF$, $\bG$ with WF $\phi$.
\end{defn}
Paying homage to the Theorem 1.1 in \cite{SY}, with similar methodology the following assertion holds true. We omit the proof.
\begin{thm}
Given SFs $\bF$ and $\bG$ in $[0,1]$, assume that a WF $x\in \bbR_+\mapsto \phi(x )\geq 0$ obeys
\beqq \int\limits_{\bbR_+} \phi(x)\big[\bF(x)-\bG(x)\big]\rd x \geq 0.\eeqq
Then
\beq D^{\rm w}_{\phi}(\bF\|\bG) \geq 0.\eeq
The equally occurs iff  $\diy\big[\frac{\bG}{\bF}-1\big]\phi=0$ for $\bF$-almost all $x\in \bbR_+$.
\end{thm}
\vskip .5 truecm

\begin{thm}
{\rm{(Estimating the WCRE via a uniform distribution, cf. Theorem 1.2 in \cite{SY}.)}}\;Assume that RV $X$ takes at most $m$ values and set $p_i=\bbP (X=i)$ and $\overline{p}_i=\sum\limits_{j=1}^i p_j$. Suppose that for given $0<\beta\leq 1$
\beqq \sum_{i=1}^m \phi(i)\big[\overline{p}_i-\beta i\big]\geq 0,\eeqq
Then
\beqq -\sum_{i=1}^m \phi(i) \overline{p}_i \log\;\overline{p}_i\leq -\log\;\beta \sum_{i=1}^m \phi(i) \overline{p}_i -\sum_{i=1}^m \phi(i) \overline{p}_i \log\;i,\eeqq
with equality iff for all $i=1\dots m$, $\phi(i)\big[\overline{p}_i-\beta i\big]=0$.\\
Furthermore, assume for given $\alpha,\beta\in \bbR$:
\beqq \int_{\bbR_+} \phi(x) \big[\bF(x)-(\alpha-\beta x)\big]\rd x \geq 0,\eeqq
The following assertion for non-negative RV $X$ holds true:
\beqq \ew(X)\leq -\int_{\bbR_+}\phi(x)\bF(x) \log (\alpha-\beta x)\rd x.\eeqq
Here the equality holds iff  $\phi(x) \big[\bF(x)-(\alpha-\beta x)\big]=0$.
\end{thm}
\vskip .5 truecm
\begin{defn}
Consider a random vector $\BX=(X_1,\dots,X_n): \Om\to\bbR^n$ with join survival function $\bF(\bx)=\bbP[X_1>x_1,\dots,X_n>x_n]$. The WCRE and WCE for given WF $\phi$, are defined by
\beq\label{Multi.WCRE.WCE} \begin{array}{c}
\ew(\BX)=-\diy\int\limits_{\bbR_+^n}\phi(\bx)P[|\BX|>\bx]\log P[|\BX|>\bx]\rd \bx,\\
\cew(\BX)=-\diy\int \limits_{\bbR_+^n}\phi(\bx)P[|\BX|\leq\bx]\log P[|\BX|\leq\bx]\rd \bx.\end{array}\eeq
Here $\bx=(x_1,\dots,x_n)$ and $\bbR_+^n=\big(x_i\in\bbR^n, x_i\geq 0\big)$.\\
Let $(X_1,X_2)\in \bbR^2\mapsto\phi (x_1,x_2)$ be a given bivariate WF. The {\bf conditional} WCRE of $X_1$ given $X_2$ is defined by
\beq\label{conWCRE}\begin{array}{l}\ew(X_1|X_2)\\
\quad=-\diy\int_{\bbR_+^2} \phi(x_1,x_2)\bbP(|X_1|> x_1,|X_2|> x_2)\log \frac{\bbP(|X_1|> x_1,|X_2|> x_2)}{\bbP(|X_2|>x_2)}\;\rd x_1\; \rd x_2.\end{array}\eeq
and the {\bf mutual} WCRE between non-negative random vectors $\BX$ with joint SF $\bF$ and marginal $\bF_i$, $i=1,\dots,n$ by
\beq \begin{array}{l}\tau^{\rm w}_{\phi}(\BX):=\diy D^{\rm w}_{\phi}(\bF\|\bF_1\otimes\dots\otimes\bF_n)\\
\quad=\diy\int\limits_{\bbR_+^n}\phi(\bx)\bF(\bx) \log \frac{\bF(\bx)}{\bF_1(x_1)\dots \bF_n(x_n)}\; \rd \bx. \end{array}\eeq
\end{defn}
\vskip .5 truecm
\begin{lem}
{\rm{(Bounding on conditional WCRE, Cf. Lemma 1.1 in \cite{SY}.)}}\;Let $\BX_1^2=(X_1,X_2)$ be a pair of RVs with the joint SF $\bF(x_1,x_2)$ and marginal SFs $\bF_1(x_1)$, $\bF_2(x_2)$. Suppose that the WF $\phi$ obeys
\beq \int\limits_{\bbR_+^2}\phi(\bx_1^2)\bF(\bx_1^2)\big[\bF(x_1|x_2)-1\big]\rd \bx_1^2 \leq 0\eeq
Then
\beq \ew(\BX_1^2)\leq \mathcal{E}^{\rm w}_{\psi_2}(X_2) \;\hbox{ or, equivalently, }\; \ew(X_1|X_2)\geq 0.\eeq
Here $\psi_2=\diy\int\limits_{\bbR_+}\phi(\bx_1^2)\bF(x_1|x_2)\rd x_1$ and the equality holds true iff $\phi(\bx_1^2)\big[\bF(x_1|x_2)-1\big]=0$ for all $\bx\in \bbR_+^2$.\\
Furthermore, consider triple RVs $\BX_1^3=(X_1,X_2,X_3)$ and assume that
\beq \int\limits_{\bbR_+^3}\phi(\bx_1^3)\bF(\bx_1^3)\big[\bF(x_1|\bx_2^3)-1\big]\rd \bx_1^3.\eeq
Then
\beq\label{Buond.1} \ew(\BX_1^3)\leq \mathcal{E}^{\rm w}_{\psi_{23}}(\BX_2^3)\;\hbox{ or, equivalently, }\; \ew(X_1|\BX_2^3)\geq 0.\eeq
Here $\psi_{23}=\diy\int\limits_{\bbR_+}\phi(\bx_1^3)\bF(x_1|\bx_2^3)\rd x_1$.
In (\ref{Buond.1}) the equality holds true iff $\phi(\bx_1^3)\big[\bF(x_1|\bx_2^3)-1\big]=0$ for all $\bx_1^3\in \bbR_+^3$.
\end{lem}
\vskip .5 truecm
\def\E{\mathcal{E}^{\rm w}}
By subscribing $\bF$ in $f$ in Theorem 1.3 in \cite{SY} with the same arguments, the following assertion, omitting the proof, is achieved.
\begin{thm}
{\rm{(Sub-additivity of the WCRE, Cf. Theorem 1.3 in \cite{SY}.)}}\;Let $\BX_1^2=(X_1,X_2)$ be a pair of RVs with join SF $\bF(\bx_1^2)$ and marginal survival function $\bF_1(x_1)$, $\bF_2(x_2)$. Moreover suppose that the WF $(x_1,x_2)\in \bbR_+^2 \mapsto \phi(x_1,x_2)\geq 0$ obeys
\beq \label{eqthm3.1}\int\limits_{\bbR_+^2}\phi(\bx_1^2)\big[\bF(\bx_1^2)-\bF_1(x_1)\bF_2(x_2)\big]\rd \bx_1^2\geq 0.\eeq
Then
\beq \begin{array}{c} \ew(\BX_1^2)\leq
\E_{\psi _1}(X_1)+\E_{\psi _2}(X_2),\;\hbox{ or, equivalently, }\;\ew (X_1|X_2)\leq
\E_{\psi_1^{\,}}(X_1),\\
\;\hbox{ or, equivalently, }\;\tau^{\rm w}_\phi (X_1:X_2)\geq 0.\end{array}\eeq
The equality occurs iff $X_1$, $X_2$ are independent modulo $\phi$ i.e. $\phi(\bx_1^2)\bigg[1-\diy\frac{\bF_1(x_1)\bF_2(x_2)}{\bF(\bx_1^2)}\bigg]=0$ for all $\bx_1^2\in \bbR_+^2$. Here $\psi_1$ and $\psi_2$ are emerging from conditional survival functions:
\beq\psi_i=\diy \int\limits_{\bbR_+} \phi(\bx_1^2)\bF(x_j|x_i)\rd x_j,\;\;\; i,j=1,2,\;\; i\neq j.\eeq
\end{thm}
For given WF $\BX_1^3\in\bbR_+^3\mapsto \phi(\bx_1^2)\geq 0$, define
\beq\begin{array}{cl} \psi_{12}(\bx_1^2)=\diy \int\limits_{\bbR_+}\phi(\bx_1^3)\bF(x_3|\bx_1^2)\rd x_3,\;\;\; \bx_1^2\in \bbR_+^2.\\
\psi_1^{23}(x_1)=\diy \int\limits_{\bbR_+^2} \phi(\bx_1^3)\bF(\bx_2^3|x_1)\rd \bx_2^3,\;\;\; x_1\in \bbR_+.\end{array}\eeq
and similarly define $\psi_k^{ij}$ and $\psi_{ij}$ for distinct labels $1\leq i,j,k\leq 3$. Then as in \cite{SY}, use $\psi_{12}$ in (\ref{eqthm3.1}) if the assumption
\beq \int\limits_{\bbR_+^3}\phi(\bx_1^3)\big[\bF(\bx_1^2)-\bF_1(x_1)\bF_2(x_2)\big]\bF(x_3|\bx_1^2) \rd \bx_1^2\geq 0.\eeq
holds true, Then
\beq  \diy \mathcal{E}^{\rm w}_{\psi_{12}}(X_1|X_2)\leq \mathcal{E}^{\rm w}_{\psi_1^{23}}(X_1).\eeq

Following the given assertions in \cite{SY}, the analogue inequalities each requiring its own assumption are represented. Note that in the list of assumptions (1.15),(1.17),(1,22),(1.27) in \cite{SY} swap $\bF$ in $f$:
\beq\label{eq:1.27}\begin{array}{l}\hbox{by Lemma 1.1, \cite{SY}:}\;\;0\leq\;\ew (X_1|\BX_2^3),
\quad\hbox{assuming 1.17 (a modified form of 1.15),}\\
\;\;\diy\begin{array}{ll}
\hbox{by Lemma 1.3, \cite{SY}:}&\ew (X_1|\BX_2^3)\leq \mathcal{E}^{\rm w}_{\psi_{12}}(X_1|X_2), \;
\hbox{assuming (1.27),}\\
\hbox{by Theorem 1.3, \cite{SY}:}&\mathcal{E}^{\rm w}_{\psi_{12}}(X_1|X_2)\leq \mathcal{E}^{\rm w}_{\psi_1^{23}}(X_1),
 \;\hbox{assuming (1.22),}\\
 \hbox{by Lemma 1.2, \cite{SY}:}&\mathcal{E}^{\rm w}_{\psi_{12}^{\,}} (X_1|X_2)\leq \mathcal{E}^{\rm w}_\phi(\BX_{1,3}|X_2), \;
 \hbox{assuming (1.26),}\\
\hbox{by Theorem 1.4, \cite{SY}:} &\mathcal{E}^{\rm w}_\phi(\BX_{13}|X_2)\leq \mathcal{E}^{\rm w}_{\psi_{12}^{\,}}(X_1|X_2)
+\mathcal{E}^{\rm w}_{\psi_{23}^{\,}}(X_3|X_2),\;\hbox{assuming (1.27),}\\
\hbox{by Theorem 1.5, \cite{SY}:} &\ew(\BX_1^3)+\mathcal{E}^{\rm w}_{\psi_2^{13}}(X_2)\leq \mathcal{E}^{\rm w}_{\psi_{12}^{\,}}(\BX_1^2)+
\mathcal{E}^{\rm w}_{\psi_{23}^{\,}}(\BX_2^3), \; \hbox{assuming (1.27).}\end{array}\end{array}\eeq
\vskip 0.5 truecm
Next we represent a number of results which are analogue assertions in \cite{SY}, hence proofs omitted.
\begin{thm}
{\rm (Strong sub-additivity of the WCRE).} Given a triple of RVs $\BX_1^3=(X_1,X_2,X_3)$, assume that
\beq\label{assum.St.sub}\diy\int\limits_{\bbR_+^3}\phi(\bx_1^3)\Big[\bF(\bx_1^3)-\bF(x_2)\prod\limits_{i=1,3} \bF(x_i|x_2)\Big]\rd \bx_1^3 \geq 0.\eeq
is fulfilled. Then
\beq \ew(\BX_1^3)-\mathcal{E}^{\rm w}_{\psi_{2}^{13}}(X_2)\leq \mathcal{E}^{\rm w}_{\psi_{12}}(\BX_1^2)+\mathcal{E}^{\rm w}_{\psi_{23}}(\BX_2^3).\eeq
The equality holds iff RVs $X_1$ and $X_3$ are conditionally independent $X_2$.
\end{thm}
\begin{thm}
{\rm (a) (Convexity of relative WCRE).} Given a WF $x\in\bbR_+\mapsto\phi (x)$ and $\lambda_1\lambda_2 \in(0,1)$ with $\lambda_1+\lambda_2=1$, then
\beq \lambda_1 \Dwphi(\bF_1\|\bG_1)+\lambda_2\Dwphi(\bF_2\|\bG_2)\geq \Dwphi(\lambda_1\bF_1+
\lambda_2 \bF_2\|\lambda_1\bG_1+\lambda_2\bG_2),\eeq
with equality iff $\lam_1\lam_2 =0$ or $\bF_1=\bF_2$ and $\bG_1=\bG_2$ modulo $\phi$.

{\rm{(b)}} {\rm{(Data-processing inequality for relative WCRE).}} Let $(\bF,\bG)$ be the pair of SFs and $\phi$ a WF in $\bbR_+$. For given stochastic kernel ${\mbox{\boldmath${\Pi}$}}=(\Pi (x,y), x,y\in\bbR_+)$, set $\Psi(u)=\diy\int_{\bbR_+} \phi(x)\Pi (u,x)\rd x$.
Then
\beq\label{eq:2.3}D^{\rm{w}}_{\Psi} (\bF||\bG)\geq \Dwphi(\bF{\mbox{\boldmath${\Pi}$}}\,\|\,\bG{\mbox{\boldmath${\Pi}$}})\eeq
where $\big(\bF{\mbox{\boldmath${\Pi}$}}\big)(x)=\diy\int_{\bbR_+}\bF(u)\Pi (u,x)\rd u$ and
$\big(\bG{\mbox{\boldmath${\Pi}$}}\big)(x)=\diy\int_{\bbR_+}\bG(u)\Pi (u,x)\rd u$. The equality occurs iff
$\bF{\mbox{\boldmath${\Pi}$}}=\bF$ and $\bG{\mbox{\boldmath${\Pi}$}}=\bG$.
\end{thm}
\begin{thm}
Let $\BX_1^3$ be a triple of  RVs with joint SF $\bF(\bx_1^3)$. Let $\bx_1^3=(x_1,x_2,x_3)\in\bbR_+^3\mapsto \phi(\bx_1^3)$ be a WF such that $X_1$ and $X_3$ are conditionally independent given $X_2$ modulo $\phi$.\\

{\rm{(a)}} {\rm (Data-processing inequality for conditional WCRE).} Assume inequality \eqref{assum.St.sub} by swapping $X_2$ with $X_1$. Then
the following assertion for conditional WCREs holds true:
\beq \label{eq:2.5}\mathcal{E}^{\rm w}_{\psi_{23}}(X_3|X_2) \leq \mathcal{E}^{\rm w}_{\psi_{13}}(X_3|X_1),\eeq
with equality iff $X_2$ and $X_3$ are independent modulo $\phi$. In addition assume that given WF $\phi$ and triple of RVs $\BX_1^3$ obey
\beq \int_{\bbR_+^3}\phi(\bx_1^3)\bF(\bx_1^3)\Big[\bF_{2|13}(x_2|\bx_{13}-1\Big]\rd \bx_1^3\leq 0\eeq
Then
\beq \mathcal{E}^{\rm w}_{\psi_{13}}(X_3|X_1)\leq 2 \mathcal{E}^{\rm w}_{\psi_{23}}(X_3|X_2);\eeq
{\rm{(b)}} {\rm (Data-processing inequality for mutual WCRE).}
Assume inequality \eqref{eqWCR:2.8}:
\beq\label{eqWCR:2.8}
\diy\int_{\bbR_+^3}\phi (\bx_1^3)\Big[\bF(\bx_1^3)-\bF_3(x_3)\prod\limits_{i=1,2}\bF_{i|3}(x_i|x_3)
\Big]\rd\bx_1^3\geq 0\eeq
Then
\beq\label{eqWCR:2.9}\tau^{\rm w}_{\psi_{13}^{\,}}(X_1:X_3) \leq \tau^{\rm w}_{\psi_{12}^{\,}}(X_1:X_2).\eeq
Here, equality in \eqref{eqWCR:2.9} holds iff, modulo $\phi$, RVs $X_1$ and $X_2$ are conditionally
independent given $X_3$.
\end{thm}

\begin{thm}\label{thm:concavity.WCRE}
{\rm (Concavity of the WCRE).} Given WF $\phi$, set $\phi'(x)=\diy\frac{\rd}{\rd x}\phi(x)$ and $\phi''(x)=\diy \frac{\rd^2}{\rd x^2}\phi(x)$. The functional $\bF\mapsto \ew(1-\bF)$ is concave function in $\bF$ under following suppositions:\\
\\
{\rm (i)} The WF $\phi$ is non-increasing (non-decreasing) for $x\in[e^{-1},1]\;(x\in [0,e^{-1}])$.\\
{\rm (ii)} For $x\in[0,1]$
\beq \diy \phi''(\bF^{-1}(x))-\diy\frac{f'(\bF^{-1}(x))}{f(\bF^{-1}(x))} \phi'(\bF^{-1}(x)) \leq 0,\eeq
here $f'$ denotes the derivative of $f$ w.r.t. $x$.
\end{thm}

{\bf Proof.}\; Set $g(x)=x\log x,\; x\in[0,1]$. To implement the concavity property for WCRE, it is sufficient to prove the function $\phi(\bF^{-1}(x)).g(x)$, $x\in[0,1]$ is convex. Therefore we compute
\beq \diy\frac{\rd}{\rd x} \phi(\bF^{-1}(x)).g(x)=\diy \big(\frac{\rd}{\rd x} \phi(\bF^{-1}(x)). g(x)+\phi(\bF^{-1}(x)).\frac{\rd}{\rd x} g(x).\eeq
and
\beq \label{eq.second.deriv}\begin{array}{l} \diy\frac{\rd^2}{\rd x^2} \phi(\bF^{-1}(x)).g(x)\\
\quad=\diy \big(\frac{\rd^2}{\rd x^2}\phi(\bF^{-1}(x))\big).g(x)+\big(\frac{\rd^2}{\rd x^2} g(x)\big). \phi(\bF^{-1}(x))+2\frac{\rd}{\rd x}g(x). \frac{\rd}{\rd x}\phi(\bF^{-1}(x)).\end{array}\eeq
Evidently the middle expression in RHS of above inequality is non-negative. Furthermore, note that
\beq\label{eq.1}\begin{array}{cl} \diy\frac{\rd}{\rd x}\phi(\bF^{-1}(x))=-\diy\frac{\phi'(\bF^{-1}(x))}{f(\bF^{-1}(x))},\\
\diy \frac{\rd^2}{\rd x^2}\phi(\bF^{-1}(x))=\frac{1}{f^2(\bF^{-1}(x))}\Big[\phi''(\bF^{-1}(x))-\diy\frac{f'(\bF^{-1}(x))}{f(\bF^{-1}(x))} \phi'(\bF^{-1}(x))\Big].\end{array}\eeq
Combining (\ref{eq.second.deriv}) and (\ref{eq.1}), under assumptions (i) and (ii) we conclude the result. $\quad$ $\Box$

\section{Additional results}

Following steps in the proof of Theorem 1 from \cite{RCVW}, we propose the following theorem.
\begin{thm}\label{Finity.WCRE}
Assume for given $0<a<\infty$ the following integrals are finite:
\beq\label{cond:1}\diy \int\limits_{(0,a)^n}\phi(\bx)\rd \bx <\infty\;\; \textrm{and}\;\; \diy\int\limits_{\bbR_+^n/(0,a)^n} \phi(\bx) \prod\limits_{i=1}^n x_i^{\frac{-p\alpha}{n}}\rd \bx<\infty\eeq
Then $\ew(\BX)<\infty$ if for all $i$, $p$ and some $0\leq \alpha \leq 1$, $\bbE[X_i^p]<\;\infty$.

Furthermore, set $\psi(x)=\diy\int_0^x\phi(t)\rd t$, in particular, $\phi(\bx)=\prod\limits_{i=1}^n \phi(x_i)$. Then for all $a>0$ the assumptions (\ref{cond:1}) take the form:
\beqq \diy \int_a^\infty \phi(x_i)\; x_i^{\frac{-p\alpha}{n}}\rd x_i <\; \infty,\;\;\; \psi(a)-\psi(0)<\infty.\eeqq
\end{thm}

{\bf Proof.}\;Following arguments given in \cite{RCVW}, we using H\"{o}lder's inequality. Recall Step 2 in the Theorem 1 of \cite{RCVW}. For $0\leq \alpha \leq 1$ we have
\beqq P[X_i>x_i,1\leq i \leq n]|\log P[X_i>x_i,1\leq i \leq n]| \leq \frac{e^{-1}}{1-\alpha}\prod\limits_{i=1}^n P[X_i>x_i]^{\frac{\alpha}{n}}.\eeqq
By multiplying both sides of above inequality in $\phi(\bx)$ and then integrating on $\bbR_+^n$, we obtain
\beqq \ew(\BX)\leq \frac{e^{-1}}{1-\alpha} \int\limits_{\bbR_+^n}\phi(\bx) \prod\limits_{i=1}^n P[X_i>x_i]^{\frac{\alpha}{n}} \rd \bx.\eeqq
Furthermore,
\beq \ew(\BX)\label{eqq:1} \leq \frac{e^{-1}}{1-\alpha} \Big[\int\limits_{(0,a)^n} \phi(\bx)\rd \bx +\int\limits_{\bbR_+^n/(0,a)^n} \phi(\bx) \prod\limits_{i=1}^n P[X_i>x_i]^{\frac{\alpha}{n}} \rd \bx\Big].\eeq
Owing to Markov inequality for $p\geq 0$, the last term of RHS in (\ref{eqq:1}) is less and equal than:
\beqq \frac{e^{-1}}{1-\alpha}\bigg[ \prod\limits_{i=1}^n \bbE[X_i^p]^{\frac{\alpha}{n}}\bigg]\int_{\bbR_+^n/(0,a)^n}\phi(\bx)\prod\limits_{i=1}^n x_i^{-\frac{p\alpha}{n}}\rd \bx.\eeqq
By virtue of (\ref{cond:1}) this leads directly to the result. $\quad$ $\Box$\\
\vskip .5 truecm
{\bf Remark.}\;Note that in case $\phi(\bx)=\prod\limits_{i=1}^n \phi(x_i)$, the (\ref{eqq:1}) reads
\beqq \begin{array}{l}
 \ew(X)\leq \diy\frac{e^{-1}}{1-\alpha}\prod\limits_{i=1}^n\int_0^\infty \phi(x_i) P[X_i>x_i]^{\frac{\alpha}{n}}\rd x_i\\
 \quad\leq \diy\frac{e^{-1}}{1-\alpha}\prod\limits_{i=1}^n\bigg\{ (\psi(a)-\psi(0))+\bbE[X_i^p]^{\frac{\alpha}{n}}\int_a^\infty \phi(x_i) x_i^{\frac{-p\alpha}{n}}\rd x_i\bigg\}.
 \end{array}\eeqq
\vskip .5 truecm
Using the method in Theorem 5, \cite{R} and arguments in Theorem \ref{Finity.WCRE}, the following result is given. The proof of Theorem \ref{thm:convergence} is similar to Theorem 5 in \cite{R} and omitted.
\begin{thm}\label{thm:convergence}
{\rm{(Cf. Theorem 5 from \cite{R}.)}}\; Let the random vector $\BX_k$ converges in distribution to the random vector $\BX$. Also suppose that $\phi$ is a WF whereas (\ref{cond:1}) holds true. If all $\BX_k$ are bounded in $L^p$ then
\beq \lim\limits_{k} \ew(\BX_k)=\ew(\BX).\eeq
\end{thm}
\vskip .5 truecm
Now we focus on the sum of independent RVs: The standard Shannon and cumulative entropies of a sum of independent RVs is larger than and equal of each . We show as analogues as Theorem 2 in \cite{RCVW}, the same result is fulfilled for WCRE either.
\begin{thm}
Consider two non-negative and independent RVs $X$ and $Y$ with PDFs $f_X$ and $f_Y$, respectively. Then
\beqq \max\bigg\{\mathcal{E}^{\rm w}_{\psi_Y}(X),\mathcal{E}^{\rm w}_{\psi_X}(Y)\bigg\} \leq \ew (X+Y).\eeqq
Here $\psi_Y(x)=\diy\int f_Y(y) \phi(x+y)\rd y$ and swap $X$ with $Y$ in $\psi_X$.
\end{thm}

{\bf Proof.}\; We again follow the argument from \cite{RCVW}. By using Jensen's inequality, write:
\beq\label{eq:09} \begin{array}{l} P[X+Y>w]\log P[X+Y>w]\\
\qquad\leq\diy \int f_Y(y)P[X>w-y]\log P[X>w-y]\rd y.\end{array}\eeq
Multiply both sides by $\phi(w)$ and then integrate with respect to $w$ from $0$ to $\infty$:
\beqq\begin{array}{l}
-\ew(X+Y) \leq \diy\int f_Y(y) \int_0^\infty \phi(w)P[X>w-y]\log P[X>w-y]\rd w \rd y\\
=\diy  \int f_Y(y) \int_y^\infty \phi(w)P[X>w-y]\log P[X>w-y]\rd w \rd y\\
=\diy  \int f_Y(y) \int_0^\infty \phi(w+y)P[X>w]\log P[X>w]\rd w \rd y\\
=\diy \int_0^\infty \bigg[\int \phi(w+y)f_Y(y)\rd y\bigg] P[X>w] \log P[X>w]\rd w=-\mathcal{E}^{\rm w}_{\psi_Y}(X).\end{array}\eeqq
The first equality here is obtained because $X$ is a non-negative RV. Consequently, for $w<y$, $P[X>w-y]=1$. $\quad$ $\Box$
\vskip .5 truecm
\vskip .5 truecm
In addition, the following extended assertion of Theorem 4 from \cite{RCVW} holds true (and is straightforward).
\begin{thm}
For given independent RV $X_i$, we have
\beq \ew(\BX)=\diy\sum\limits_{i} \mathcal{E}^{\rm w}_{\phi^*_i}(X_i),\eeq
where $\phi^*_i=\diy\int \phi(\bx)\prod\limits_{j\ne i}\bF(x_j)\rd \bx_1^{i-1}\rd \bx_{i+1}^n$. In a particular case $\phi(\bx)=\prod\limits_{j=1}^n \phi_j(x_j)$, set $\psi_j(x_j)=\diy\int_0^{x_j} \phi_j(t)\rd t$. Then $\ew(\BX)$ in (\ref{Multi.WCRE.WCE}) becomes:
\beqq \ew(\BX)= \sum\limits_{i=1}^n \Bigg(\prod\limits_{j=1,j\neq i}^n \bigg[\bbE_{X_j}(\psi_j(X_j))-\psi_j(0)\bigg]\Bigg) \mathcal{E}^{\rm w}_{\phi_i}(X_i). \eeqq
\end{thm}

\section{Bounds for the WCRE}
In this section, our goal is to establish additional bounds for the WCRE. First, let us show how the WCRE can be dominated by the standard entropy, as well as by the CRE; cf. \cite{RCVW}.
\begin{thm}\label{thm:Bound1}
Let $h(X)$ be the Shannon entropy of a non-negative RV $X$ having PDF $f$ and SF $\bF$. Then
\beq\label{eq:09} \ew(X) \geq \alpha_\phi\; \exp\{h(X)\},\eeq
where
\beq\label{eq:2.03}\alpha_\phi=\diy \exp\bigg\{\bbE[\log \phi(X)]+\int_0^1\log \bigg(x|\log x|\bigg)\rd x\bigg\}.\eeq
\end{thm}
{\bf Proof.}\; The proof follows directly from log-sum inequality:
\beqq\begin{array}{l}
\diy\int f(x)\log \frac{f(x)}{\phi(x) \bF(x)|\log \bF(x)|}\rd x\\
\quad \geq \diy \Big(\int f(x) \rd x\Big)\log\Bigg(\frac{\int f(x) \rd x}{\int \phi(x)\bF(x)|\log \bF(x)|\rd x}\Bigg)=\log \frac{1}{\ew(X)}.
\end{array}\eeqq
Note that if $\diy \int \phi(x) \bF(x) \log \bF(x) \rd x=\infty$, the proof is trivial. The LHS leads:
$$-h(X)-\bbE[\log \phi(X)]-\diy \int_0^1 \log x|\log x|\rd x.$$
Consequently,
$$ \log \ew(X)\geq h(X)+\bbE[\log \phi(X)]+\diy \int_0^1 \log x|\log x|\rd x.$$
This completes the proof. $\quad$ $\Box$\\

Let $(X,Y)$ be a pair of RVs with a conditional SF $\bF(x|y)$. Moreover assume an additional WF $(x,y)\in \bbR_+^2\mapsto\varphi(x,y)\geq 0$. The WCRE of RV $X$ given $Y=y$ with WF $\varphi(x,y)$ is defined by
\beq\label{CWCRE} \mathcal{E}^{\rm w}_{\varphi}(X|Y=y)=-\int \varphi(x,y) \bF(x|y)\log \bF(x|y)\;\rd x. \eeq
Later, owing to (\ref{CWCRE}), the generalized statement of Theorem \ref{thm:Bound1} with similar proof is driven, therefore we omit the proof.

\begin{lem}
For a non-negative RV $X$, let $f(x|y)$ be the conditional PDF $X$ given $Y=y$. Set
\beqq \alpha_\varphi(y)=\diy \exp\bigg\{\bbE_{X|Y=y}[\log \varphi(X,y)]+\int_0^1\log \bigg(x|\log x|\bigg)\rd x\bigg\}\eeqq
Then
\beq\label{eq:2.02} \mathcal{E}^{\rm w}_\varphi(X|Y=y) \geq \alpha_\varphi(y)\; \exp\{h(X|Y=y)\}.\eeq
where $h(X|Y=y)$ is Shannon entropy of $X$ given $Y=y$.
\end{lem}
\vskip .5 truecm
\begin{defn}
Given WF $(x,y)\in \bbR_+^2\mapsto\varphi(x,y)\geq 0$, set $\overline{\phi}(x)=\diy \int_0^\infty \varphi(x,y) f(y|x)\rd y$. The {\bf Cross} WCRE is introduced by
\beq \mathcal{E}^{\rm w}_{\varphi}(X,Y)=\mathcal{E}^{\rm w}_{\overline{\phi}}(X)+\bbE_X\big[\mathcal{E}^{\rm w}_{\varphi}(Y|X=x)\big].\eeq
\end{defn}
\vskip .5 truecm
\begin{lem}
Assume WF $(x,y)\in \bbR_+^2\mapsto\varphi(x,y)\geq 1$ and set
\beq\label{eq:21.03}\alpha^*_\varphi=\diy \exp\bigg\{\bbE_{X,Y}[\log \phi(X,Y)]+\int_0^1\log \bigg(x|\log x|\bigg)\rd x\bigg\}.\eeq
Then
\beqq \ew(X,Y)\geq 2\alpha^*_\varphi \exp\{\frac{h(X,Y)}{2}\}.\eeqq
\end{lem}

{\bf Proof.}\;Following the assumption $\varphi\geq 1$, we observe
\beq\label{eq:22:02} \bbE_X\big[\log \overline{\phi}(X)\big]\geq \bbE_{X,Y}\big[\log \varphi(X,Y)\big].\eeq
Owing to the convexity of $e^x$ and Jensen inequality, we have
\beq\label{eq:2.04} \bbE\big[\mathcal{E}^{\rm w}_{\varphi}(Y|X=x)\big]\geq \alpha^*_\varphi \exp\{h(Y|X)\}.\eeq
where $h(Y|X)$ is denoted for conditional standard entropy.\\
\beq\label{eq22:01} \begin{array}{ccl}
\mathcal{E}^{\rm w}_{\varphi}(X,Y)&=&\mathcal{E}^{\rm w}_{\overline{\phi}}(X)+\bbE_X\big[\mathcal{E}^{\rm w}_{\varphi}(Y|X=x)\big]\\
&\geq& \alpha_{\overline{\phi}}\exp\{h(X)\}+\alpha^*_{\varphi}\exp\{h(Y|X)\}\\
&\geq& 2\alpha^*_{\varphi}\exp\{\frac{h(X,Y)}{2}\}.\end{array}\eeq
Here $\alpha_{\overline{\phi}}$ is defined as (\ref{eq:2.03}) by replacing $\overline{\phi}$ in $\phi$. The first inequality in (\ref{eq22:01}) drives from (\ref{eq:09}) and (\ref{eq:2.04}). The second inequality holds by using (\ref{eq:22:02}) and $2\exp(\diy\frac{t+s}{2})\leq \exp(t)+\exp(s)$.  $\quad$ $\Box$
\vskip .5 truecm
\begin{lem}
{\rm{(Cf. Proposition 4 from \cite{RCVW}.)}} Let $X$ be a non-negative continuous RV. Given WF $\phi$, suppose that $\psi(x)=\diy\int_0^x\phi(t)\rd t$ is bounded. There exist a function $Y=g(X)$ such that:
\begin{itemize}
\item[] (i) The WCRE and the weighted entropy (WE) are related by
\beqq h^{\rm w}_{\phi}(Y)=\frac{\ew(X)}{\bbE(X)}+\frac{\bbE(\psi(X))-\psi(0)}{\bbE(X)}\log\;\bbE(X),\eeqq
\item[] (ii) Assume $\psi(0)=0$, then the WCRE and the Shannon entropy (SE) are related by
\beqq h(Y)=\frac{\ew(X)}{\bbE(\psi(X))}-\frac{\Theta}{\bbE(\psi(X))}+\log\; \bbE(\psi(X)),\eeqq
here $\Theta=\diy\int_0^\infty \phi(x) \log \phi(x)\; \bF(x)\rd x$.
\end{itemize}
\end{lem}

{\bf Proof.}\; The proof is straightforward by considering the CDF, $F$, as an RV having PDF $\diy\frac{P(X>x)}{\bbE(X)}$ and $\diy\frac{\phi(x)P(X>x)}{\bbE(\psi(X))}$, respectively. Next use the definitions of SE and WE.\\
Note that simply by choosing $g(x)=F^{-1}(F(x))$ we can find a $g$. $\quad$ $\Box$
\vskip .5 truecm
Next we present a Lower bound for WCR, the origin of this Lemma goes back to Proposition 1 from \cite{R}.
\begin{lem}\label{lemma2.4}
Let $X$ and $Y$ be two iid RVs. Also For given WF $\phi$ set $\psi(x)=\diy \int_0^x \phi(t)\rd t$. We obtain
\beq\label{eq:3.02} 2 \ew(X) \geq \bbE\big[|\psi(X)-\psi(Y)|\big].\eeq
In particular, suppose that $X$ is a non-negative RV, then
\beq\label{eq:3.03} 2 \ew(X) \geq \bbE\big[|\psi(X)-\bbE[\psi(X)]|\big].\eeq
\end{lem}

{\bf Proof.}\;According to Proposition 1, \cite{R}, similarly we derive:
\beq\label{eq:3.01} 2\bF(x)-2\bF^2(x)= \bbP\big[\max\{X,Y\}>x\big]-\bbP\big[\min\{X,Y\}>x\big],\eeq
multiplying both sides of (\ref{eq:3.01}) in $\phi(x)$ and then integrating from zero to infinity:
\beqq \begin{array}{l} 2 \diy\int_0^\infty \phi(x) \bF(x)(1-\bF(x))\rd x\\
\quad=\diy\int_0^\infty \phi(x) \bbP\big[\max\{X,Y\}>x\big]- \diy\int_0^\infty \phi(x) \bbP\big[\min\{X,Y\}>x\big].\end{array}\eeqq
Next using integrate by part in RHS , we can write
\beq\label{eq:Lemma 2.4}2\int_0^\infty \bF(x)|\log \bF(x)|\rd x \geq \bbE\big[|\psi(X)-\psi(Y)|\big].\eeq
The LHS can be modified becaue of $x(1-x)\leq x|\log x|$.
The inequality (\ref{eq:Lemma 2.4}) proves (\ref{eq:3.02}). Moreover the assertion (\ref{eq:3.03}) follows directly from:
\beqq \bbE\big[|\psi(X)-\psi(Y)|\big]\geq \bbE\big[|\psi(X)-\bbE[\psi(X)]|\big].\qquad \Box \eeqq
\vskip .5 truecm
\begin{lem}
{\rm{(Cf. Proposition 2 from \cite{R}.)}} Let $X$ be a non-negative RV. Then for function $\psi$ defined as in Lemma \ref{lemma2.4}:
\beq \ew(X)=\bbE\bigg[(\psi(0)-\psi(X))(1+\log \bF(X))\bigg].\eeq
\end{lem}

{\bf Proof.}\;The proof is straightforward and based on the equality:
\beqq \bF(x)\log \bF(x)=-\int_x^\infty (1+\log \bF(t))\rd \bF(t).\qquad \Box\eeqq
\vskip .5 truecm
{\bf Remark:}\; More application of conjugate or the Fenchel Transform of the convex function $x\log x$ is $\exp(y-1)$, that is
\beqq \exp(y-1)=\sup\left[xy-x\log x:0<x<\infty\right].\eeqq
Consequently, for non-negative RVs $X$ and $Y$:
\beqq xy\leq x\log x +\exp(y-1).\eeqq
If we use this inequality, emerging the definition WCRE, an upper bound for WCRE in terms of $|\psi(X)-\bbE[\psi(X)]|$ is given:
\beqq \ew(X) \leq 2\bbE\bigg[|\psi(X)-\bbE[\psi(X)]|\log |\psi(X)-\bbE[\psi(X)]|\bigg]+\frac{4}{e}.\eeqq
Here $\psi$ is defined as before.
\vskip .5 truecm
\begin{thm}
{\rm{(Cf. Theorem 1 from \cite{R}.)}}\;Suppose that $X$ is a non-negative RV. Set $\psi(x)=\diy\int_0^x \phi(t)\rd t$ and $\psi^{-1}$ the inverse function of $\psi$. Then
\beq \label{maineq:2.1} \begin{array}{l}\bbE\big[\psi(X)\log^+\psi(x)\big]\\
\qquad \diy\leq \ew(X)+\bbE\big[\psi(X){\mathbf 1}(X>\psi^{-1}(1))\big]\log \bigg(e\;\bbE\big[\psi(X){\mathbf 1}(X>\psi^{-1}(1))\big]\bigg).\end{array}\eeq
This implies: $\bbE\big[\psi(X)\log^+\psi(x)\big]<\infty$ if WCRE is finite.
\end{thm}

{\bf Proof.}\;Following standard calculations, (see \cite{R}), we can write
\beq \label{eq:2.05} \begin{array}{l}\bbE\big[\psi(X)\log^+\psi(x)\big]\\
\qquad=\diy\bbE\big[\psi(X){\mathbf 1}(X>\psi^{-1}(1))\big]-\bF(\psi^{-1}(1))+\diy\int_{\psi^{-1}(1)}^\infty \phi(x)\bF(x) \log \psi(x)\rd x. \end{array}\eeq
Moreover, for $t>\psi^{-1}(1)$ one yields:
\beqq\psi(t)\bbP(X>t)\leq \bbE\big[\psi(X){\mathbf 1}(X>t)\big]\leq \bbE\big[\psi(X){\mathbf 1}(X>\psi^{-1}(1))\big].\eeqq
Therefore, we obtain
\beqq \begin{array}{l} \diy \int_{\psi^{-1}(1)}^\infty \phi(x)\bF(x) \log \psi(x)\rd x\\
\quad \leq \diy \ew(X)+\log \bbE\big[\psi(X){\mathbf 1}(X>\psi^{-1}(1))\big]\diy \int_{\psi^{-1}(1)}^\infty \phi(x)\bF(x)\rd x.\end{array}\eeqq
Finally according to (\ref{eq:2.05}), we get
\beqq \bbE\big[\psi(X)\log^+\psi(x)\big] \leq \ew(X)+\varsigma\Bigg(1+ \log \bbE\big[\psi(X){\mathbf 1}(X>\psi^{-1}(1))\big]\Bigg).\eeqq
Here $\varsigma=\diy\bbE\big[\psi(X){\mathbf 1}(X>\psi^{-1}(1))\big]-\bF(\psi^{-1}(1))$. The inequality (\ref{maineq:2.1}) holds true then. $\quad$ $\Box$
\vskip .5 truecm


\section{Maximum WCRE properties}

\begin{thm}\label{thm3:1}
Suppose $x\in \bbR_+\mapsto \phi(x)\geq 0$ is given WF. Then $\bF^{\rm m}$ maximizes the WCRE $\ew(\bF)$, modulo $\phi$, uniquely when the following constrains are fulfilled:
\beq \int\limits_{\bbR_+}\phi(x)\big[\bF(x)-\bF^{\rm m}(x)\big]\rd x\geq 0 \quad \textrm{and} \quad \int\limits_{\bbR_+}\phi(x)\big[\bF(x)-\bF^{\rm m}(x)\big]\log \bF^{\rm m}(x) \rd x\geq 0.\eeq
\end{thm}
\vskip .5 truecm
\def\bt{\mathbf{t}} \def\bmu{\mathbf{\mu}} \def\BC{\mathbf{C}}
\begin{exam}\label{exam:max.G}
Consider a random vector $\BX_1^n=(X_1,\ldots ,X_n):\Om\to\bbR^n$ with PDF $f$, the PDF $F$ and the SF $\bF$, mean vector $\bmu =(\mu_1,\ldots ,\mu_n)$ with
 $\bbE X_i=\mu_i$ and covariance matrix $\BC=(C_{ij})$ with $C_{ij}=\bbE\Big[(X_i-\mu_i)(X_j-\mu_j)\Big]$,
$1\leq i,j\leq n$.  Let $f^{\rm{No}}$ be the normal PDF with the same  ${\mbox{\boldmath${\mu}$}}$ and $\BC$ and $\bF^{\rm{No}}$ be the normal SF. Introduce
\beq \label{defn.alpha*}\alpha^*(\bx)=\diy\int\limits_{\bx}^{\infty} \exp\Big\{-\frac{1}{2}(\bt-\bmu)^T \BC^{-1}(\bt-\bmu)\Big\} \rd\bt.\eeq
Then
\beq \label{CDF.MNormal} \diy\rho(\BC):=\bF^{\rm{No}}(\bx)=\diy(2\pi)^{-n/2}({\rm det}\; \BC)^{-1/2} \alpha^*(\bx). \eeq
Given a WF $\bx_1^n=(x_1,\ldots ,x_n)\in\bbR^n\mapsto\phi (\bx_1^n)\geq 0$, suppose that
\beq \begin{array}{l}
\diy\int\limits_{\bbR_+^n}\phi(\bx)\Big[\bF(\bx)-\bF^{\rm{No}}(\bx)\Big] \rd \bx \geq 0\;\; \hbox{and}\\
\diy \log\big[(2\pi)^{n/2} ({\rm det}\;\BC)^{1/2}\big]\int\limits_{\bbR_+^n} \phi(\bx)\big[\bF(\bx)-\bF^{\rm No}(\bx)\big]\rd \bx -
\int\limits_{\bbR_+^n} \phi(\bx)\big[\bF(\bx)-\bF^{\rm No}(\bx)\big] \log \alpha^*(\bx)  \rd \bx \leq 0.\end{array}\eeq
Then
\beq \begin{array}{l}\ew(F)\leq \ew(F^{\rm No})\\
\quad=\diy \frac{1}{2} \log \big[(2\pi)^n ({\rm det}\;\BC)\big]\int\limits_{\bbR_+^n}\phi(\bx)\bF^{\rm No}(\bx)\rd \bx
-\int\limits_{\bbR_+^n}\phi(\bx)\bF^{\rm No}(\bx) \log \alpha^*(\bx) \rd\bx.\end{array}\eeq
with equality iff $\bF=\bF^{\rm{No}}$ modulo $\phi$.
\end{exam}
\vskip .5 truecm
\vskip .5 truecm
\begin{exam}
Let $F^{Exp}$ and $\bF^{Exp}$ be respectively CDF and SF on $\bbR_{+}$ with mean $\frac{1}{\lambda}$. Suppose the following constrains are fulfilled:
\beq\diy\int\limits_{\bbR_+}\phi(x)\bigg[\bF(x)-\bF^{Exp}(x)\bigg]\rd x\geq 0 \;\; \textrm{and}\;\; \diy\int\limits_{\bbR_+}x\;\phi(x)\bigg[\bF(x)-\bF^{Exp}(x)\bigg]\rd x\leq 0,\eeq
where $x\in \bbR_+\mapsto\phi(x)\geq0$ is a given WF. Then
\beq \ew(F)\leq \ew(F^{Exp})=\lambda \diy\int\limits_{\bbR_+} x\;\phi(x) e^{-\lambda x} \rd x.\eeq
and $\bF^{Exp}$ is a unique maximizer modulo $\phi$.
\end{exam}

The next Theorem is a direct result of Theorem \ref{thm:concavity.WCRE} and Example \ref{exam:max.G}.
\begin{thm}
{\rm (The weighted Ky Fan inequality using the WCRE; cf. \cite{SY}, Theorem 3.2).}\;Assume for given $\lambda_1,\lambda_2\in[0,1]$ with $\lambda_1+\lambda_2=1$ and positive definite matrices $\BC_1$, $\BC_2$ and $\BC=\BC_1+\BC_2$ the assumption in Theorem \ref{thm:concavity.WCRE}, (i) and (ii) hold true. Furthermore
\beq\begin{array}{l}
\diy\int\limits_{\bbR_+^n}\phi(\bx)\big[\lambda_1 \bF_{\BC_1}^0(\bx)+\lambda_2\bF_{\BC_2}^0(\bx)-\bF_{\BC}^0(\bx)\big]\rd \bx\geq 0\;\;\; \hbox{and}\\
\diy \log\big[(2\pi)^{n/2} ({\rm det}\;\BC)^{1/2}\big]\int\limits_{\bbR_+^n} \phi(\bx)\big[\lambda_1 \bF_{\BC_1}^0(\bx)+\lambda_2\bF_{\BC_2}^0(\bx)-\bF_{\BC}^0(\bx)\big]\rd \bx \\
\diy-\int\limits_{\bbR_+^n} \phi(\bx)\big[\lambda_1 \bF_{\BC_1}^0(\bx)+\lambda_2\bF_{\BC_2}^0(\bx)-\bF_{\BC}^0(\bx)\big] \log \alpha_{\BC}^*(\bx)  \rd \bx \leq 0.\end{array}\eeq
are fulfilled. Then
\beq \rho(\lambda_1\BC_1+\lambda_2\BC_2)-\lambda_1\rho(\BC_1)-\lambda_2\rho(\BC_2)\geq 0.\eeq
with equality iff $\lambda_1\lambda_2=0$ or $\BC_1=\BC_2$.
\end{thm}

\vskip .5 truecm

\begin{lem}\label{lem4.1}
Let $\BX_1^n=(X_1,\ldots ,X_n)$ be a random vector, with components $X_i:\Om\to\cX_i$, $1\leq i\leq n$,
and the joint SF $\bF$. Introduce the random vector $\overline{\bx}_i=(\bx_1^{i-1},\bx_{i+1}^n)$, $\bF_i(x_i)$ the marginal SF for RV $X_i$:
\beqq \bF_i(x_i)=\lim _{\overline{\bx}_i\rightarrow \infty} \bF(\bx)\;\;\; \hbox{and}\;\;\; \bF_{|i}(\bx_1^n|x_i)=\diy\frac{\bF(\bx_1^n)}{\bF_i(x_i)}.\eeqq
For given a WF $\phi$, suppose that
\beq \diy \int\limits_{\bbR_+^n} \phi(\bx)\Big[\bF(\bx)-\prod\limits_{i=1}^n \bF_i(x_i)\Big]\rd\bx \geq 0.\eeq
Then
\beq\label{Ineq.M} \ew(\BX)\leq \sum\limits_{i=1}^n \mathcal{E}^{\rm w}_{\psi_i} (X_i).\eeq
where
\beq \psi_i(x_i)=\int \limits_{\bbR_+^{n-1}}\phi(\bx_1^n)\bF_{|i}(\bx_1^n|x_i)\rd \bx_1^{i-1}\rd \bx _{i+1}^n.\eeq
The equality in (\ref{Ineq.M}) holds true holds iff, modulo $\phi$, components $X_1, 1dots, X_n$ are independent.
\end{lem}
In the following theorem a straightforward application of Lemma \ref{lem4.1} is given.
\begin{thm}
{\rm (The weighted Hadamard inequality using the WCRE; cf. \cite{SY}, Theorem 3.3).}\;Let  $\BC=(C_{ij})$ be a positive definite $n\times n$ matrix
and $\bF^{\rm{No}}_{\BC}$ the normal SF with the zero mean vector and the covariance matrix $\BC$. For given WF $\bx_1^n=(x_1,\ldots ,x_n)\in\bbR^n\mapsto\phi (\bx_1^n)$, introduce $\alpha^*(\bx)$ by (\ref{defn.alpha*}) and
\beq \alpha^*_i(x)=\diy\int\limits_{x}^\infty e^{-t^2\big/2C_{ii}}\rd t\;\;\;\; \hbox{and}\;\;\;\; \alpha=\diy \int\limits_{\bbR_+^n} \phi(\bx) \bF^{\rm{No}}(\bx)\rd \bx.\eeq
Suppose that
\beq \diy \int\limits_{\bbR_+^n} \phi(\bx)\Big[\bF^{\rm {No}}(\bx)-\prod\limits_{i=1}^n \bF^{\rm {No}}_i(x_i)\Big]\rd\bx \geq 0.\eeq
Then
\beq \diy \frac{\alpha}{2}\log \bigg[\prod\limits_i C_{ii}\Big/ {\rm det}\;\BC\bigg]+\int\limits_{\bbR_+^n} \phi(\bx) \bF^{\rm{No}}(\bx) \log \bigg[\alpha^*(\bx)\Big/\prod\limits_i \alpha^*_i(x_i)\bigg] \rd \bx \geq 0,\eeq
with equality iff $\BC$ is diagonal.
\end{thm}

\vskip .5 truecm
Next, we provide a characterization of the Weibull distribution using the maximum WCRE.
\begin{thm}
{\rm{(Cf. Theorem 2 from \cite{R}.)}}\;Suppose $\psi^*_p(x)=\diy\int_0^x t^p\phi(t)\rd t$ is a non-negative WF such that $\psi(x)=\diy\int_0^x \phi(t)\rd t$ and $x\in \bbR^+\mapsto \phi(x)\in[0,1]$. Among all non-negative RVs with given $\bbE[\psi(X)]$ and $\bbE[\psi^*_p(X)]$ the Weibull distribution $W$ with SF $\bF_{\rm Wib}(t)=\exp(-\lambda^q t^q)$, has the maximal WCRE.\\
Here the parameters $q=p$ and
\beq\label{lambda.Weibull} \lambda^q=\bigg(\frac{c_p}{\bbE[\psi(X)]-\psi(0)}\bigg)^p.\bigg(\frac{\bbE[\psi^*_p(X)]-\psi^*_p(0)}{\bbE_{\rm Wib}[\psi^*_q(X)]-\psi^*_q(0)}\bigg);\eeq
where $c_p=\Gamma(1+\frac{1}{p})$.
\end{thm}

{\bf Proof.}\; The subsequence argument works by using Log-sum inequality once more. According to (21) in \cite{R} but replacing $\bG(x)=\bF_{\rm Wib}(x)=\exp(-\mu^p x^p)$ in (\ref{eq:4.01}), we get
\beqq \ew(X) \leq \bigg[\bbE[\psi(X)]-\psi(0)\bigg] \log \frac{\bbE[\psi(X)]-\psi(0)}{\mu^{-1}c_p} +\mu^p \int_0^\infty \phi(t) t^p \bF(t) \rd t.\eeqq
Now choose $\mu^{-1}c_p=\bbE[\psi(X)]-\psi(0)$:
\beqq \ew(X)\leq \frac{c_p^p \big(\bbE[\psi^*_p(X)]-\psi^*_p(0)\big)}{\bbE[\psi(X)]-\psi(0)}.\eeqq
Finally let $q=p$ and $\lambda$ as in (\ref{lambda.Weibull}), therefore we have
\beqq \ew(X) \leq \lambda^q \bigg(\bbE_{\rm Wib}[\psi^*_q(X)]-\psi^*_q(0)\bigg)=\ew(Wib).\eeqq
This completes the proof. $\quad$ $\Box$

\vskip .5 truecm
\begin{thm}
Suppose that functions $\psi$ and $\phi$ are as in Theorem 3.2: $\psi (x)=\diy\int_0^x \phi(t)\rd t)$, and $0\leq\phi (x)\leq 1$.
Let $X$ be a given non-negative RV. In addition assume $Z:=X(\lambda)$ is an exponentially distributed RV with mean
$\lambda^{-1}=\bbE[\psi(X)]-\psi(0)$. If the constraints
\beq\label{assum3.1} \diy\int\limits_{\bbR_+}x\; \phi(x) \big[\bF(x)-\bF^{\rm Exp}(x)\big]\rd x \geq 0.\eeq
holds true, then
\beq\label{maxi.Exp} \ew(X)\leq \ew(X(\lambda))= \lambda\diy\int\limits_{\bbR_+} x\;\phi(x)\bF^{\rm Exp}(x) \rd x.\eeq
\end{thm}

{\bf Proof.}\; Using log-sum inequality we obtain
\beq\label{eq3:1} \begin{array}{l}
\diy\int\limits_{\bbR_+} \phi(x) \bF(x) \log \big(\phi(x)\bF(x) e^{\lambda\;x}\big)\rd x\\
\qquad \geq \diy\bigg(\int\limits_{\bbR_+}\phi(x) \bF(x) \rd x\bigg)\log \bigg(\lambda \int\limits_{\bbR_+} \phi(x) \bF(x) \rd x\bigg).\end{array}\eeq
We also can write
$$\bbE[\psi(X)]-\psi(0)=\diy\int\limits_{\bbR_+}\phi(x)\bF(x)\rd x, $$
therefore the expression (\ref{eq3:1}) becomes
\beqq \begin{array}{l}
-\ew(F)+\diy\int\limits_{\bbR_+} \bF(x) \phi(x) \log \phi(x)\rd x+\lambda \int\limits_{\bbR_+}x\;\phi(x) \bF(x)\rd x\\
\qquad \geq \diy\big(\bbE[\psi(X)]-\psi(0)\big)\bigg\{\log \lambda+\log \big(\bbE[\psi(X)]-\psi(0)\big)\bigg\}.\end{array}\eeqq
Equivalently
\beq \label{eq3.2}\begin{array}{ccl}
-\ew(F)\geq \diy &-&\int\limits_{\bbR_+} \bF(x) \phi(x) \log \phi(x)\rd x-\lambda \int\limits_{\bbR_+}x\;\phi(x) \bF(x)\rd x\\
&+& \diy \big(\bbE[\psi(X)]-\psi(0)\big)\bigg\{\log \lambda+\log \big(\bbE[\psi(X)]-\psi(0)\big)\bigg\}.\end{array}\eeq
Now, set
$$\varpi=\bigg(\bbE[\psi(X)]-\psi(0)\bigg)^2 \Big(\diy\int\limits_{\bbR_+} x\;\phi(x)\bF(x) \rd x\Big)^{-1},\;\; \lambda^*=\frac{\varpi}{\bbE[\psi(X)]-\psi(0)}$$
It is admissible (\ref{eq3.2}) is fulfilled for all positive $\lambda$, so is also valid for maximum value of $\lambda=\lambda^*$. This represents the formula:
\beqq\begin{array}{ccl}
-\ew(F)\\
\quad&\geq& -\diy\int\limits_{\bbR_+} \bF(x) \phi(x) \log \phi(x)\rd x-\big[\bbE[\psi(X)]-\psi(0)\big]+\diy\big[\bbE[\psi(X)]-\psi(0)\big]\log \varpi\\
&\geq& -\diy\big[\bbE[\psi(X)]-\psi(0)\big]+\big[\bbE[\psi(X)]-\psi(0)\big]\log\varpi\\
&\geq& -\diy\big[\bbE[\psi(X)]-\psi(0)\big]+\big[\bbE[\psi(X)]-\psi(0)\big]\diy\bigg\{1-\frac{1}{\varpi}\bigg\}\\
&=& -\diy \big[\bbE[\psi(X)]-\psi(0)\big] \varpi^{-1}.
\end{array}\eeqq
The second inequality holds true owing to $\phi\in[0,1]$ and the last inequality is satisfied by using $\log x\geq 1-\frac{1}{x}$, $x\in \bbR_+$.\\
Recalling assumption (\ref{assum3.1}), leads to (\ref{maxi.Exp}). $\quad$ $\Box$
\vskip .5 truecm

{\emph{Acknowledgements --}}
YS thanks the Math Department, Penn State University,  for the financial support and hospitality
during the academic year 2014-5. SYS thanks the CAPES PNPD-UFSCAR Foundation
for the financial support in the year 2014-5. SYS thanks
the Federal University of Sao Carlos, Department of Statistics, for hospitality during the year 2014-5.


\end{document}